# Comparing Unit Trains versus Manifest Trains for the Risk of Rail Transport of Hazardous Materials - Part II: Application and Case Study

Di Kang[a], Jiaxi Zhao[b], C. Tyler Dick[b], Xiang Liu*[a], Zheyong Bian[c], Steven W. Kirkpatrick[d], Chen-Yu Lin[e]

[a] Department of Civil and Environmental Engineering, Rutgers, The State University of New Jersey, Piscataway, NJ.

[b] Rail Transportation and Engineering Center – RailTEC, Department of Civil and Environmental Engineering, University of Illinois at Urbana Champaign, Urbana, IL

[c] Department of Construction Management, University of Houston, Houston, TX.

[d] Applied Research Associates, Inc., Los Altos, CA.

[e] Department of Transportation and Logistics Management, National Yang Ming Chiao Tung University, Taipei, Taiwan.

**Corresponding Author:**

**Xiang Liu**, xiang.liu@rutgers.edu

**Abstract:** Built upon the risk analysis methodology (presented in the part I paper), this part II paper focuses on applying this methodology. Five illustrative scenarios were used to analyze the best or worst cases and compare the transportation risk differences between service options using unit trains and manifest trains. The comparison results indicate that if all tank cars are placed at the positions with the lowest probability of derailing and if switching tank cars alone in classification yards, it could provide the lowest risk estimate given the same transportation demand (i.e., number of tank cars to transport). This paper also shows that based on the data and parameters in the case study, risks during arrival/departure events and yard switching events could be as significant as risks that on mainlines. This paper provides a way to use the risk analysis methodology for rail safety decisions. The methodology and its application can be tailored to specific infrastructure and rolling stock characteristics.

## 1. Introduction

In 2020, U.S. Class I railroads[1] transported 2.1 million carloads and 175 million tons of chemicals (Association of American Railroads, 2021b). According to the Association of American Railroads (AAR) (2021a), "more than 99.99% of all hazardous material (hazmat) moved by rail arrives at its destination without a release caused by a train accident". Although rail transportation in the United States is considered as the safest way of moving large quantities of hazmat over long distances, the consequences of potential accidents and releases from a train carrying multiple hazmat cars (can be up to 120 in a unit train) is much greater than that by a traditional truck trailer carrying a single hazmat car. There have been a number of server railroad hazmat transportation accidents, causing significant impacts. For example, on 6 July 2013, the Lac-Mégantic rail tragedy resulted in 47 victims caused by the derailment and release of a 73-car unit train carrying Bakken Formation crude oil.

Railroad hazardous materials transportation generally experienced two (three) types of risks for unit trains (manifest trains): unit trains experience line-haul risks on mainline segments and arrival/departure (A/D) risks arriving at or departing from the terminal, while manifest trains incur the additional risks associated with switching railcars in classification yards. These risks are structured as multi-event, multi-factor risk management problems in part I of this paper, which are analyzed based on the event sequence, including train accident occurrence, number of cars derailed, number of hazmat cars derailed, number of hazmat cars releasing, and release consequences. Section 3 in part I of this paper proposed a methodology for a risk analysis comparison of unit and manifest trains considering both the mainline and yard/terminal risk components of freight rail transportation. To demonstrate the application of the proposed approach to an actual railway hazmat transportation scenario, this paper presents a detailed step-by-step case study calculation and comparison using the methodology outlined in Section 3 of the part I paper.

[1] As of 2019, the Surface Transportation Board defines a Class I as having operating revenues of, or exceeding, $505 million annually. (Resource: https://en.wikipedia.org/wiki/Railroad_classes.)

The case study is inspired by an actual hazmat unit train derailment and release incident. On November 7, 2013, a southbound Alabama & Gulf Coast Railway (AGR) train was traveling from Amory, Mississippi, towards Walnut Hill, Florida, with 88 loaded hazmat tank cars, two buffer cars, and three locomotives. The case study is designed to evaluate the relative risk of making a similar crude oil shipment in a single unit train as compared to multiple manifest trains. We assume that 100 high-hazard flammable tank cars need to be transported from Amory, Mississippi, to Walnut Hill, Florida (approximately 400 miles). In general, two service options are proposed to transport these 100 tank cars to compare the risks related to each operating strategy: 1) one unit train with 100 tank cars; 2) five manifest trains, each including 80 non-tank cars and a block of 20 tank cars. Regardless of the train type (unit or manifest train), the following assumptions are made for these two operating strategies:

- Train operating speed is 25 mph on mainline and 15 mph in terminals/yards.
- Each train has five locomotives (each weighing 212.5 tons).
- Each railcar is loaded with a gross railcar weight of 143 tons (regardless of car type).
- Tank cars are all DOT 117s.
- Each manifest train is routed through three classification yards per shipment.
- The unit train is a fixed consist from the origin terminal to the destination terminal.

- The base conditional probability of release for a DOT 117 tank car is 0.043 (Treichel et al., 2019) under this predefined context.
- Derailment rates and parameters used in various distributions are all estimated based on Class I railroad annual report financial data, Surface Transportation Board waybill sample data, and mainline and yard/terminal derailments from the FRA Rail Equipment Accident (REA) database for the years 1996-2018.

Five scenarios are considered in order to compare the risks associated with the operating strategies using one unit train or five manifest trains carrying 100 tank cars over 400 miles considering four different factors. *Train type* is the primary factor. Scenario 1 uses one unit train to transport all 100 tank cars, while scenarios 2-5 use five manifest trains to transport these 100 high-hazmat flammable tank cars. *Position in manifest train* is the second factor. Scenarios 2 and 3 place the block of 20 tank cars at the back of the train (i.e., positions with the lowest probability of derailment based on the position-dependent derailment probability distribution on the mainline to test the best-case scenario regarding tank car positions), while scenarios 4 and 5 place this block in the middle of the train (i.e., positions with the highest probability of derailment based on the position-dependent derailment probability distribution on the mainline to test the worst-case situation regarding tank car positions). The conditional position-dependent derailment probability distribution given a mainline train derailment will be discussed in detail in Section 2.2.1. *Yard switching approach* for the manifest train is the third factor to compare the relative effects of making different assumptions about how railcars are switched in yards. Scenarios 2 and 3 consider the block of 20 tank cars to be "switched alone," while scenarios 4 and 5 consider the block of 20 tank cars to be "switched en masse" with a group of 19 non-hazmat railcars in front (the reason of the number 19 has been explained in the part I paper). The “switched alone” approach generates a lower risk compared to “switched en masse” since it is not coupled to any non-hazmat railcars. Note that in this experiment design, the position of the tank car block in the manifest train on the mainline is correlated with the switching approach: tank cars positioned in the middle of the manifest train are “switched en masse” (worst-case scenario for manifest trains) while tank cars positioned at the back of the train are switched alone (best-case scenario for manifest trains). This correspondence reflects the practicalities of how the manifest train might be switched by backing it over a hump or switching lead upon arrival at a classification yard. *Yard type* is included as the fourth factor to compare the relative risks of hump and flat switching yards. Scenarios 2 and 4 use flat yards, while scenarios 3 and 5 use hump yards. The last three factors do not apply to scenario 1 since it involves unit train terminals instead of classification yards.

Although the methodology for line-haul risks does not include locomotives due to data limitations, in this case study, we include five locomotives for line-haul risk calculation since the mainline operations normally have a high speed and it is necessary to consider locomotives for a derailment incident. However, since train activities in yards/terminals proceed with a reduced speed and the mainline locomotives are not always included (e.g., the yard switching events are hauled by a switch engine), the risk calculation in yards/terminals does not include locomotives.

Due to the complexity of the methodology itself and various factors considered in the case study, each of the five scenarios is assigned a two- or four-character code for the unit train scenario (scenario 1) or manifest train scenarios (scenarios 2-4) to help keep track of the various factor levels associated with it. Each character designates the particular level of one of the four factors.

Specifically, **U-T** represents **U**nit in **T**erminal; **MBAF** represents **M**anifest, **B**ack of train, switched **A**lone, **F**lat yard type; **MBAH** corresponds to **M**anifest, **B**ack of train, switched **A**lone, and **H**ump yard type; **MMEF** represents **M**anifest, **M**iddle of train, switched **E**n masse, **F**lat yard type; and **MMEH** means **M**anifest, **M**iddle of train, switched **E**n masse, and **H**ump yard type. Table 1 summarizes these five scenarios and the corresponding factor levels associated with each scenario. Again, in Table 1, "back of train" and "middle of train" correspond to the positions with the lowest probability of derailing and highest probability of derailing (on the one-mile mainline segment), which will be detailed in Section 2.2.1.

**Table 1 Summary of case study scenarios.**

| Scenario and Code | Train type | Number of trains needed to transport 100 tank cars | Position of 20 tank car block in manifest train | Switching approach | Number of terminals or classification yards | Yard type |
|---|---|---|---|---|---|---|
| 1<br>U-T | Unit train | 1 | N/A | N/A | 1 origin<br>1 destination | Terminal |
| 2<br>MBAF | Manifest train | 5 | Back of train | Alone | 1 origin<br>1 intermediate<br>1 destination | Flat |
| 3<br>MBAH | Manifest train | 5 | Back of train | Alone | 1 origin<br>1 intermediate<br>1 destination | Hump |
| 4<br>MMEF | Manifest train | 5 | Middle of train | En Masse | 1 origin<br>1 intermediate<br>1 destination | Flat |
| 5<br>MMEH | Manifest train | 5 | Middle of train | En Masse | 1 origin<br>1 intermediate<br>1 destination | Hump |

Due to the large number of scenarios, we first calculate the releasing consequence (expected casualties) related to each risk component (line-haul risk, A/D risk, and yard switching risk), and then combine them for each scenario considering different factor levels accordingly. Instead of calculating the total risk associated with each scenario one by one, calculating the risks based on risk components separately could avoid unnecessary repetition for similar scenarios.

The remainder of this paper is organized as follows: in Section 2, we provide all calculation details of this case study and show the differences between using unit trains and manifest trains with certain train configurations. Insights are discussed in Section 3 based on the results from Section 2. Section 4 conducts a sensitivity analysis with various operation speed on mainlines. Section 5 concludes this case study and discusses possible future work.

Since this paper frequently refers to the part I paper (Kang et al., 2022), we do not cite it in the following context to avoid redundancy.

## 2. Detail Calculation

### *2.1 Derailment Likelihood*

#### 2.1.1 Derailments on Mainlines

Given the historical train derailment data on mainlines for the years 1996-2018, summarized in Section 3 of the part I paper, we initially categorize each cause into train-mile-based, car-mile-based, and ton-mile-based cause groups (Table A.1 in Appendix A). Assume a mainline segment $i$ has a length of one mile. According to the methodology described in Section 3.1.1 of the part I paper and the shipment information described above, the line-haul train derailment probabilities per shipment on this one-mile mainline segment $i$ for the unit train and the manifest train are 8.53E-07 and 9.54E-07, respectively (Table A.1 in Appendix A). Since transporting 100 tank cars requires one unit train or five manifest trains in the predefined context, the line-haul train derailment probabilities per traffic demand over 400 miles are $(8.53^{-07} \times 1 \times 400)$ for the unit train and $(9.54^{-07} \times 5 \times 400)$ for the manifest train.

#### 2.1.2 Derailments in Yards and Terminals

Based on the analysis of all yard/terminal derailments considering A/D events for the years 1996-2018 (Zhao & Dick, 2022), the proportion of derailments attributable to train-mile or car-mile causes for unit trains and manifest trains are shown in Table 2. Following Section 3.1.2 of the part I paper, the A/D derailment likelihoods per shipment are calculated separately for the unit train in terminals (Table 3) and the manifest train in yards (Table 4). The case study calculation of the A/D derailment likelihood combines the train-mile and car-mile (or trains processed and railcars processed) A/D derailment rates introduced in Section 3.1.2 of the part I paper to reflect derailment causes linked to each respective metric unit. It distinguishes between yard type, with separate manifest train A/D likelihood assuming all three yards are hump classification yards (Scenarios 3 and 5) or flat switching yards (Scenarios 2 and 4). For this 400-mile/three-classification-yard case study shipment using manifest trains, the hump yards yield a lower A/D derailment likelihood as compared to flat yards (the latter is almost three times larger than the former). One possible reason to explain a higher risks in flat yards than hump yards is that hump yard has dedicated tracks for arrival and departure events and separates these two switching processes, but flat yards mess arrival and departure processes up. In this analysis framework, the position of the tank cars in the middle or back of the manifest train does not influence the A/D derailment likelihood since the block of tank cars will traverse 400 miles/three classification yards no matter where they are placed. Hence, scenarios 2 and 4 and scenarios 3 and 5 have the same A/D derailment likelihood even though they involve trains with tank cars at different positions in the train. However, this factor of position in a manifest train will be important for later calculations of derailment severity.

The probability of a yard switching derailment per shipment for the manifest train is calculated in Table 5 according to Section 3.1.2 of the part I paper. In addition to yard type (hump or flat yards), the yard switching derailment likelihood calculation distinguishes the yard switching approaches ("switched alone" or "switched en masse"). The case study scenarios switching in flat yards exhibit slightly lower yard switching derailment likelihoods than hump yards, and scenarios using the "switched alone" approach have a significantly lower yard switching derailment likelihood than scenarios using the "switched en masse" approach (the latter is almost two times larger than the former).

**Table 2 Proportion of yard and terminal derailments attributed to train-mile and car-mile causes by train type (1996-2018) (Zhao & Dick, 2022)**

| Train type | Train-mile causes | Car-mile causes |
|---|---|---|
| Manifest train | 78.1% | 21.9% |
| Unit train | 62.8% | 37.2% |

**Table 3 The probability of A/D derailment for the unit train in terminals per shipment**

| Metric unit | Metric unit proportion (Table 2) | The number of trains or cars involved per A/D event | The number of A/D events involved per shipment | The number of A/D train derailments per million train A/D events (Table 3 in part I paper) | The A/D derailment probability for the unit train per shipment |
|---|---|---|---|---|---|
| Train-mile cause | 62.8% | 1 (train) | 2 | 126.31 | 2.53E-04 |
| Car-mile cause | 37.2% | 100 (cars) | 2 | 1.22 | 2.44E-04 |
| **Total** | 62.8% * 2.53E-04+37.2%*2.44E-04= | | | | **2.49E-04** |

**Table 4 The probability of A/D derailment for the manifest train in yards per shipment**

| Metric unit | Metric unit proportion (Table 2) | The number of trains or cars involved per A/D event | The number of A/D events involved per shipment | The number of A/D train derailments per million train A/D events (Table 3 in part I paper) | | The A/D derailment probability for the manifest train per shipment | |
|---|---|---|---|---|---|---|---|
| | | | | Hump yard | Flat yard | Hump yard | Flat yard |
| Train-mile cause | 78.1% | 1 (train) | 1 (at origin yard) + 2 (at intermediate | 36.53 | 118.92 | 1.46E-04 | 4.76E-04 |
| Car-mile cause | 21.9% | 100 (cars) | | 0.62 | 2.02 | 2.48E-04 | 8.08E-04 |
| | | | yard) + 1 (at destination yard) | | | | |
| **Total** | Hump yard: 78.1%*1.46E-04+21.90%*2.48E-04= | | | | | **1.68E-04** | |
| | Flat yard: 78.1%*4.76E-04+21.90%8.08E-04= | | | | | | **5.48E-04** |

**Table 5 The probability of yard switching derailment for the manifest train per shipment**

| Yard switching approach | Number of cars involved per yard switching event | Number of yard switching events per shipment | The number of yard switching derailments per million cars processed in the yard (Table 3 in part I paper) | | The probability of the yard switching derailment per shipment | |
|---|---|---|---|---|---|---|
| | | | Hump yard | Flat yard | Hump yard | Flat yard |
| Switched alone | 20 tank cars | 1 (at origin) and 1 (at intermediate yard) | 6.49 | 6.38 | **2.60E-04** | **2.55E-04** |
| Switched en masse | 19 non-tank cars and 20 tank cars | | | | **5.06E-04** | **4.98E-04** |

Finally, the probabilities of line-haul derailments, A/D derailments, and yard switching derailments (hump yard or flat yard) are summarized in Table 6. Overall, the derailment likelihood on a 1-mile mainline segment is three orders of magnitude less likely to derail as compared to an A/D derailment or yard switching derailment in terminals and yards. This is because the metric for the mainline derailment likelihood in Table 6 is “per mile per shipment” while it is “per shipment” in terminals and yards. Comparing the arrival/departure (for options with the unit train and the manifest train) and yard switching derailment likelihoods (for options with the manifest train) across all case study scenarios, the service option using the unit train or the option with manifest train routing through hump yards consistently exhibits lower yard/terminal derailment probabilities than other service options.

**Table 6 Summary of the derailment probabilities for the unit train and manifest train per shipment**

| Derailment type | Derailment location | Yard switching approach | Train type | |
|---|---|---|---|---|
| | | | Unit train | Manifest train |
| Line-haul risk (per mile per shipment) | On mainline segments | - | 8.53E-07 | 9.54E-07 |
| Arrival/departure risk (per shipment) | In terminals | - | 2.49E-04 | - |
| | In flat yards | - | - | 5.48E-04 |
| | In hump yards | - | - | 1.68E-04 |
| Yard switching risk (per shipment) | In flat yards | Switched alone | - | 2.55E-04 |
| | | Switched en masse | - | 4.98E-04 |
| | In hump yards | Switched alone | - | 2.60E-04 |
| | | Switched en masse | - | 5.06E-04 |

## 2.2 Number of Hazmat Cars Releasing Contents Per Derailment

### 2.2.1 Line-haul Incidents on Mainlines

According to Equation (11) in part I of this paper, the probability of the railcar at each position being the point of derailment during the line-haul process on segment $i$ is shown in Figure 1. Since the placement of the block of tank cars in manifest trains is a factor affecting derailing consequences, the derailment probability at each position in a manifest train is calculated following Sections 3.2.1 and 3.3.1 of the part I paper (Figure 2). As can be concluded from Figure 2, when the 20 hazmat cars are placed in the middle of the case study manifest train (train consist is shown in Figure 3(b)), they have a higher chance of derailing during the line-haul process as compared to when they are placed at the back of the train (train consist is shown in Figure 3(a)). For unit trains traversing the origin and destination terminals through mainlines, the position-dependent derailment probability is also calculated in Figure 2(c). Since the unit train is only composed of tank cars, there are no alternative railcar arrangements to consider.

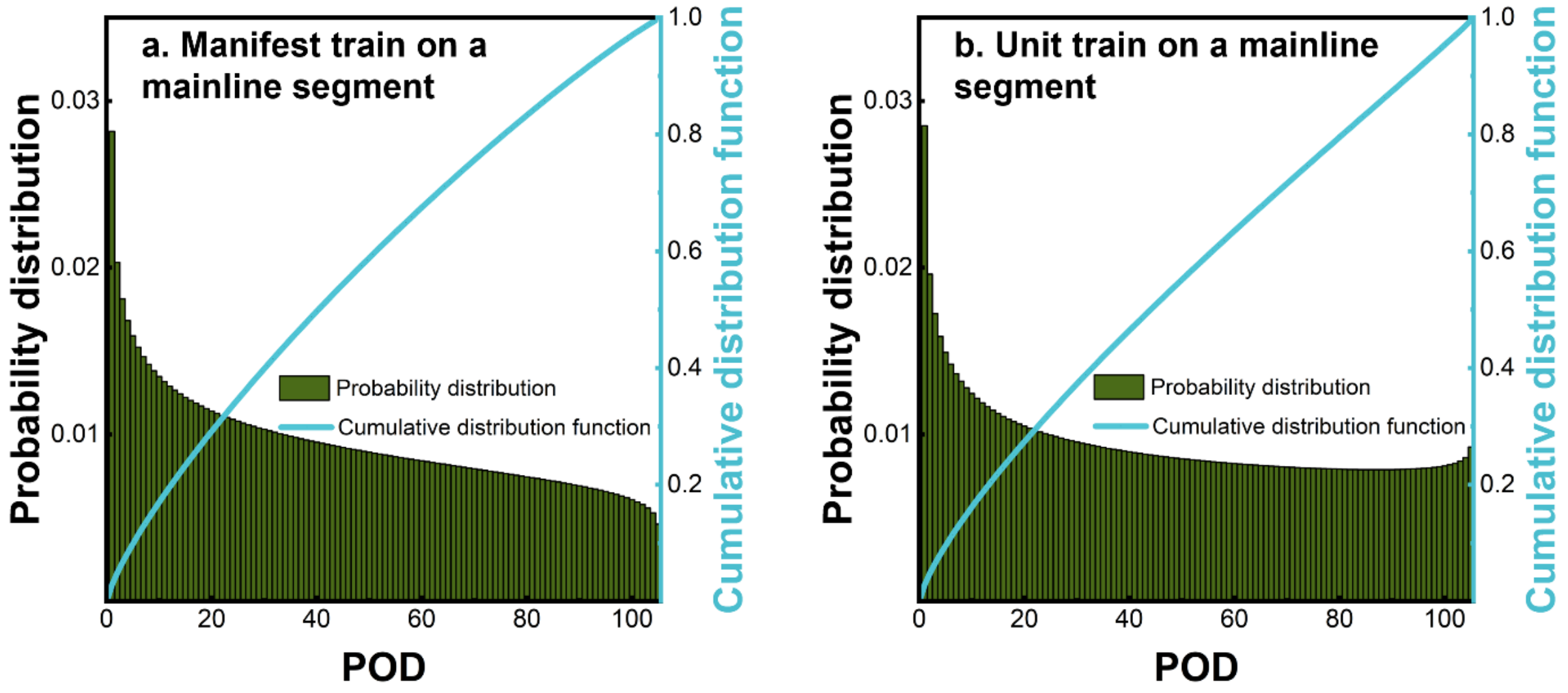

**Figure 1 Probability of railcars at each position being the point of derailment for (a) manifest train and (b) unit train on a mainline segment.**

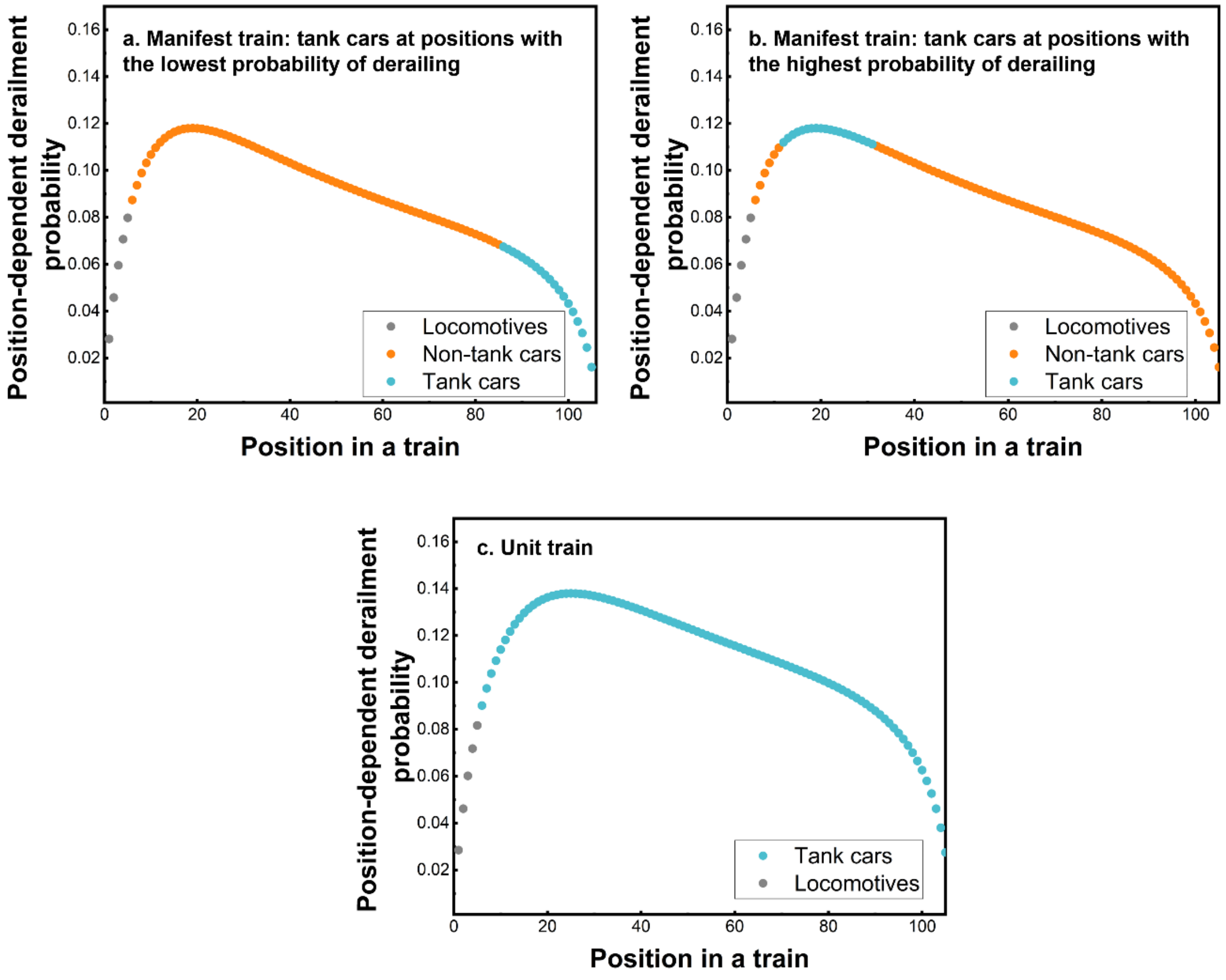

**Figure 2 Position-dependent derailment probability during line-haul transportation for (a) manifest train placing the block of tank cars at positions with the lowest derailment probability, (b) manifest train placing the block of tank cars at positions with the highest derailment probability, (c) unit train.**

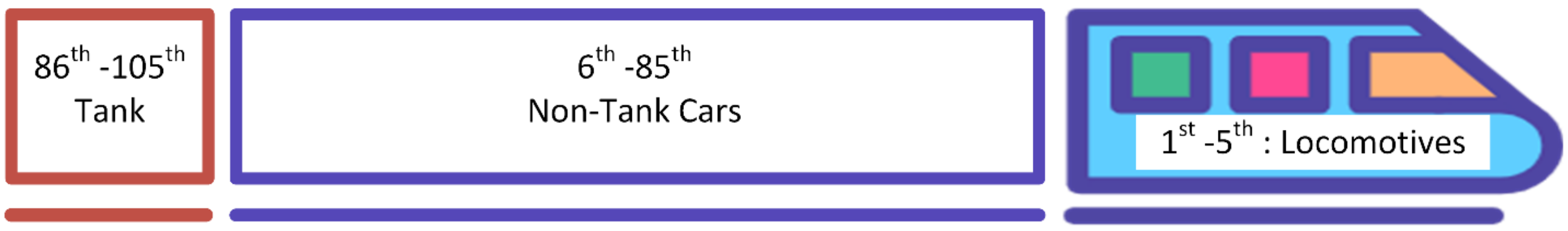

**(a) Tank cars at the back of the train**

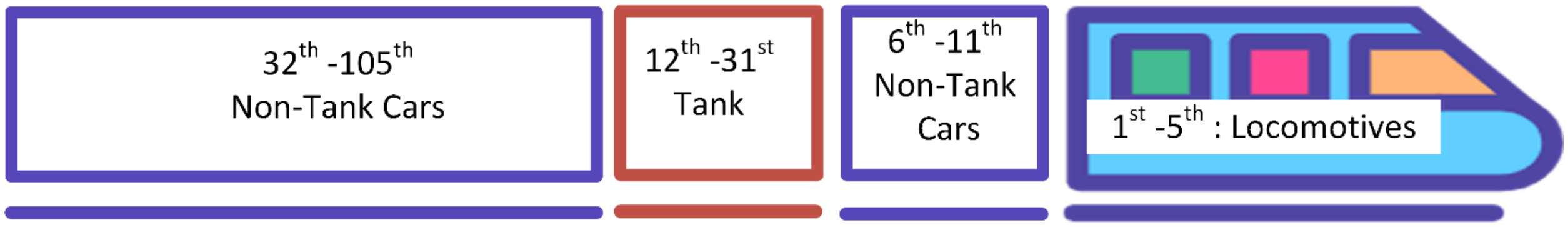

**(b) Tank cars in the middle of the train**

**Figure 3 Train consists of the five manifest trains (a) tank cars at the back of the train (scenarios 2 and 3), and (b) tank cars in the middle of the train (scenarios 4 and 5).**

According to Section 3.3.1 of the part I paper, the probability distribution of the number of hazmat cars releasing contents on a mainline segment, given a train derailment, is shown in Figure 4. The probability distributions in Figure 4 indicate that service options with manifest trains tend to have a larger probability of releasing no tank cars while the unit train has larger probabilities of releasing one to two tank cars. This originates from different train consists between unit and manifest trains. A manifest train has 80 non-tank cars that can derail with no release, which accounts for a large probability of no tank cars releasing. On the contrary, since a unit train consists of five locomotives and 100 tank cars, it has a greater opportunity to derail and result in at least one tank car releasing once a train derailment occurs on a mainline segment. Cases with more than ten tank cars releasing contents are cut in Figure 4 since they have negligible probabilities of occurring.

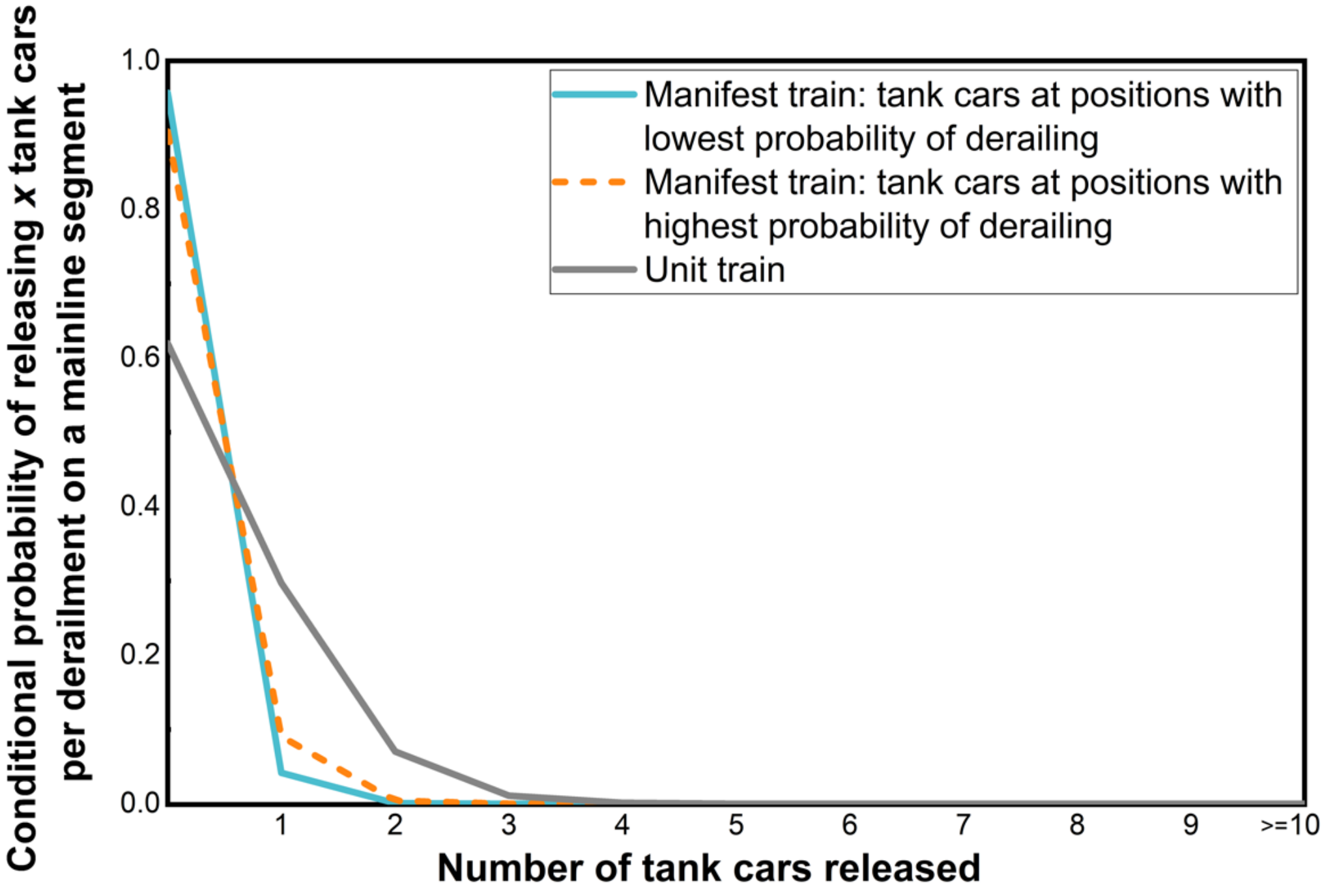

**Figure 4 Conditional probability distribution of the number of tank cars releasing contents per derailment for the line-haul incident on a 1-mile mainline segment.**

### 2.2.2 A/D Incident in Yards/Terminals

Following Section 3.2.1 of the part I paper, the probability distributions of POD at each position given an A/D event are plotted in Figure 5 for manifest trains and unit trains. These fitting results indicate that the POD in a unit train skews to the front of a train, while the manifest train has a significantly smaller probability at the first few positions as the POD. The different shapes of the two cumulative distributions in Figure 5(a) and Figure 5(b) also demonstrate the different characteristics between the unit train and manifest train pertaining to transportation risks.

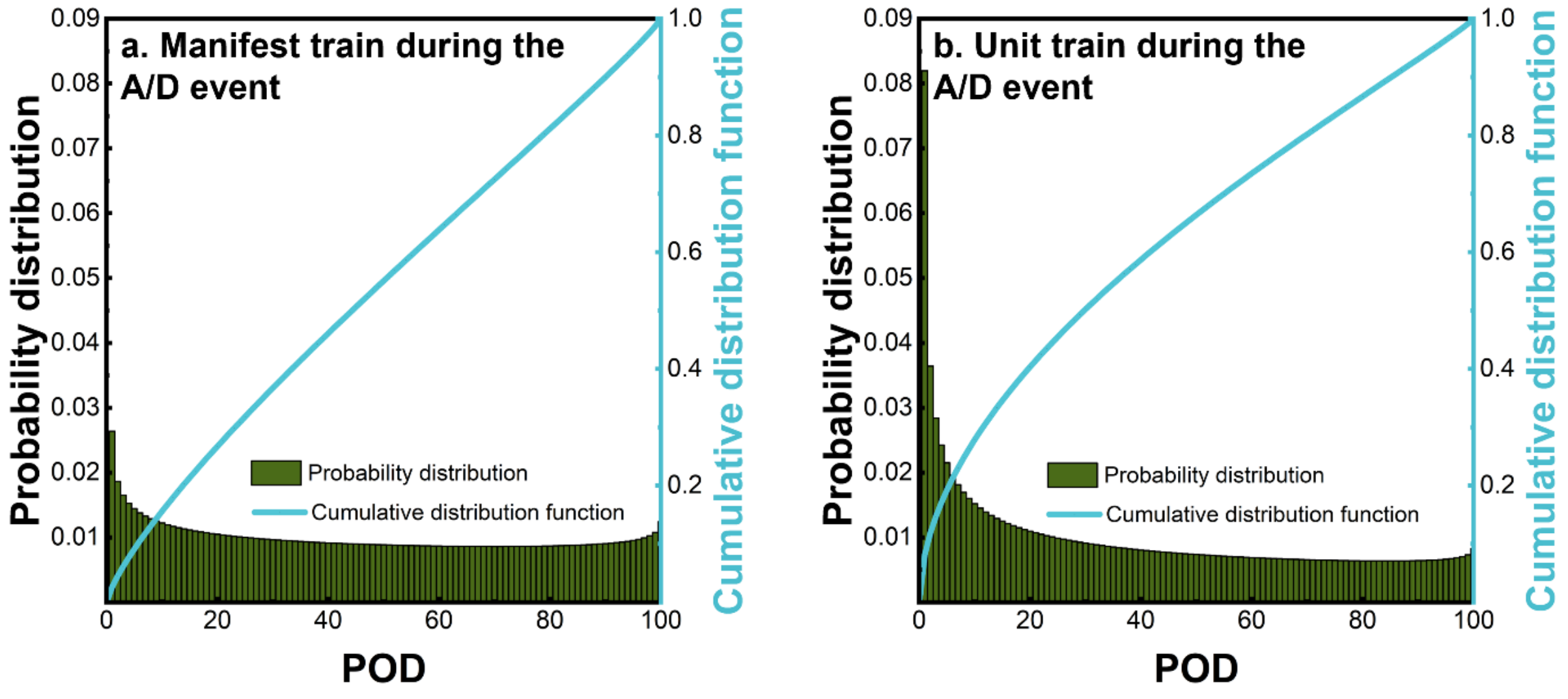

**Figure 5 Probability of railcars at each position being the point of derailment for (a) manifest train and (b) unit train during an A/D event.**

Section 3.3.2 of the part I paper proposes an approach to obtain the probability distribution of the number of tank cars derailed given an A/D train derailment (Figure 6). Since the manifest train is assumed to ship 20 tank cars along with 80 non-tank cars (train consists in yards/terminals do not consider locomotives, as described), even if a manifest train experiences an A/D derailment in the yard, there is still a possibility that the derailed railcars are all non-tank cars. Therefore, the sum of conditional probabilities over each number of tank cars derailed is less than one for manifest trains since the higher probability of derailing zero tank cars is not plotted for clarity. For the case study scenarios using manifest trains and placing tank cars at positions with the lowest chance to derail, 82.5% of the A/D derailments only involve non-tank cars (i.e., zero tank cars derailed), and it is 62.9% for scenarios using manifest trains and placing tank cars at positions with the highest probability of derailing.

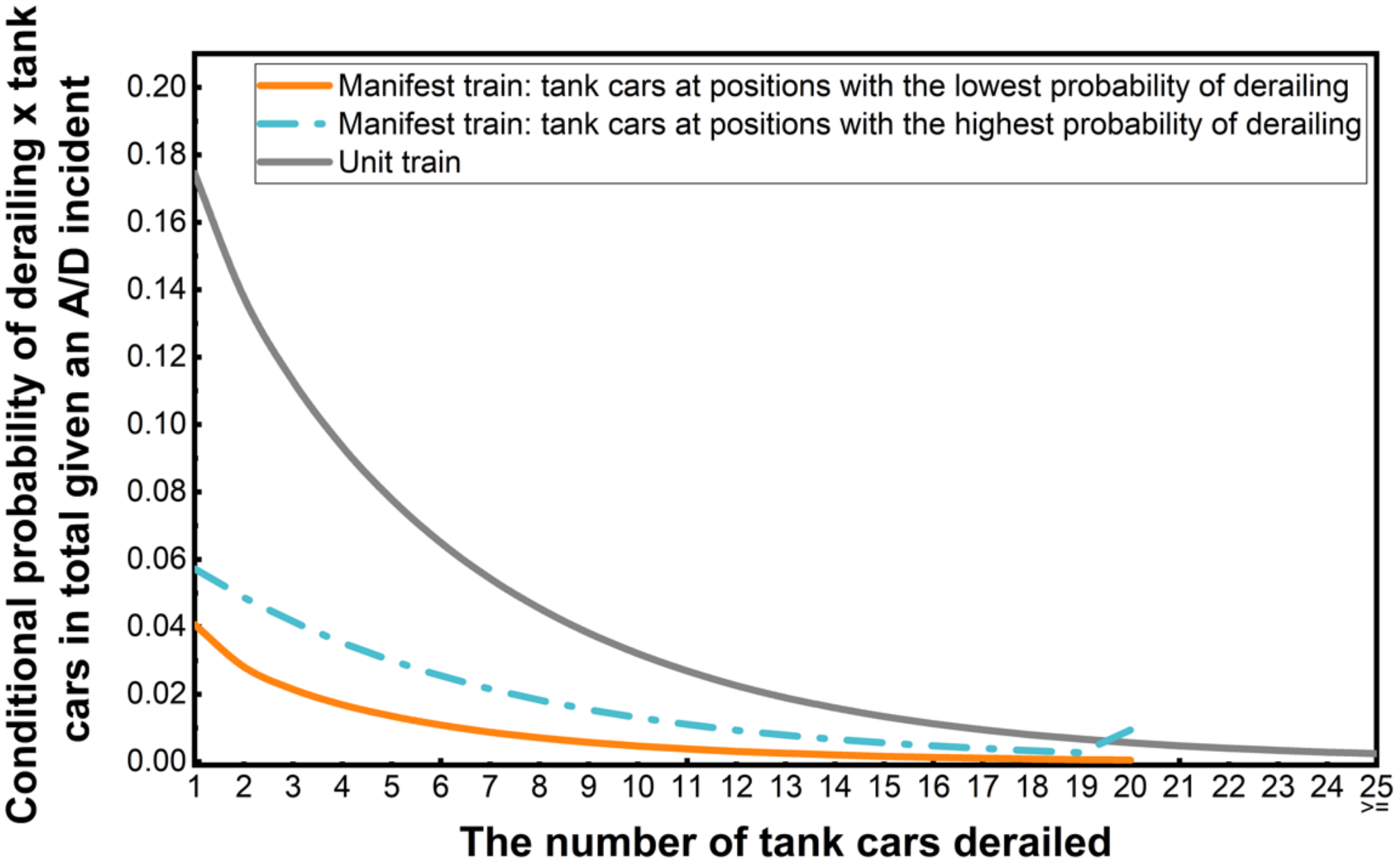

**Figure 6 Conditional probability distribution of the number of tank cars derailed given an A/D incident.**

In comparison, since the unit train only contains tank cars, the case study scenarios using unit trains involve derailing at least one tank car given an A/D incident. Hence, the sum of conditional probabilities over each number of tank cars derailed per A/D derailment is exactly one. Note that the maximum number of tank cars derailed in a manifest train is 20; and the cases that derailed more than 25 tank cars in a unit train are cut in Figure 6 since the corresponding probabilities are almost zero.

According to Equation (26) of the part I paper, the conditional probability distribution of the number of tank cars releasing contents given an A/D incident can be developed, as shown in Figure 7. Once an A/D incident occurs, the unit train has a higher likelihood of small-scale tank car release (releasing 1-2 tank cars) than manifest trains due to the larger number of tank cars (100) on the case study unit train compared to the manifest train (20 tank cars). Regardless of train type or the placement of tank cars in a manifest train, a train is less likely to release more than three tank cars during A/D events, and the probability decreases as severity increases for all scenarios.

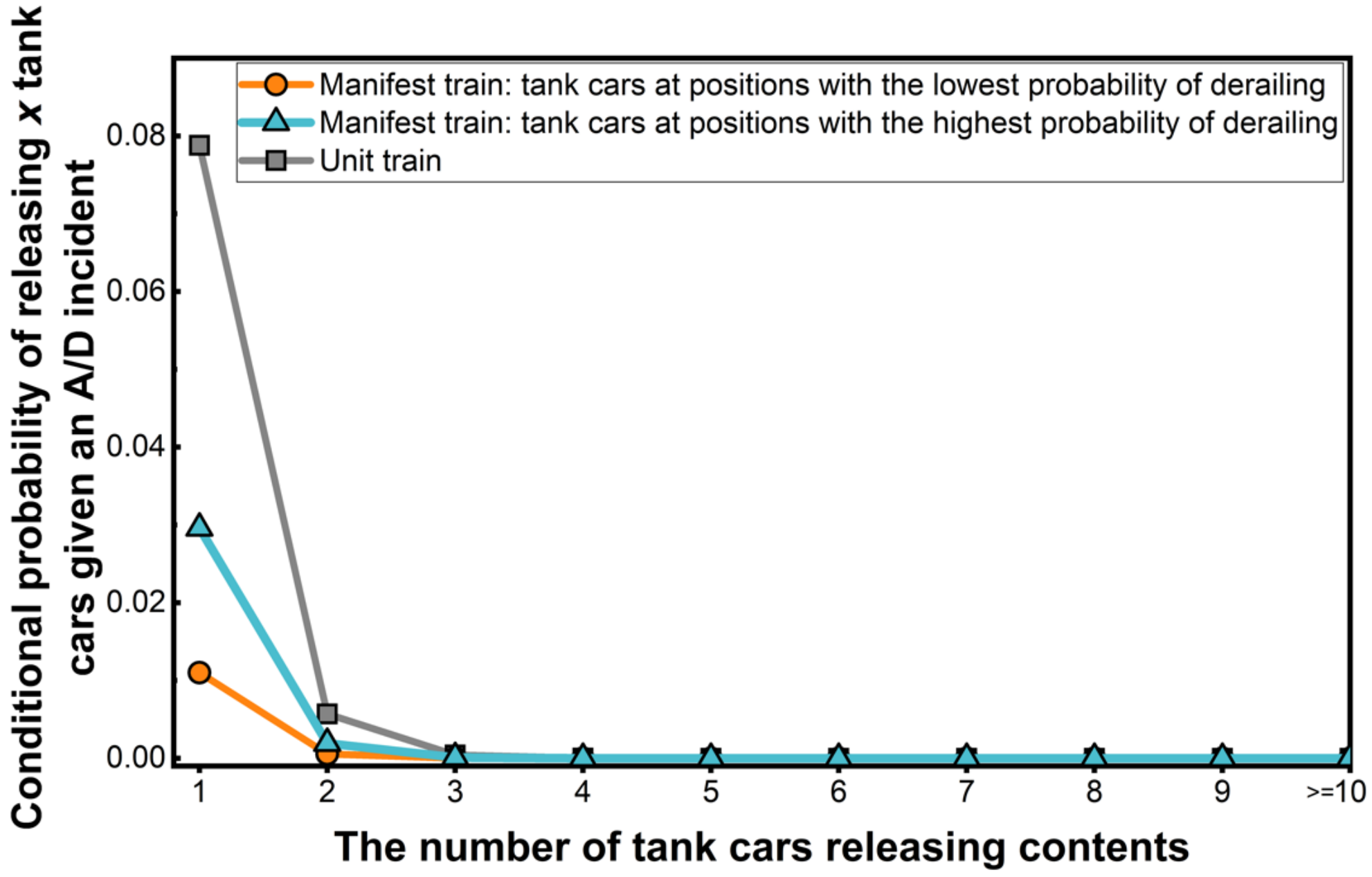

**Figure 7 Conditional probability distribution of the number of tank cars releasing contents given an A/D incident.**

### 2.2.3 Yard Switching Incidents

Given that a yard switching derailment occurs, the conditional probability distribution of derailing $x$ tank cars in a manifest train exhibits different characteristics depending on yard type (flat yard or hump yard) and yard switching approach (switched alone or switched en masse), as discussed in the part I paper. According to Section 3.2.2 of the part I paper, different yard types result in different derailment severities, as calculated by Equations (18) and (19) in part I paper for flat yards and for hump yards (Figure 8). Figure 8 indicates that derailments in flat yards, compared with those in hump yards, tend to have smaller probabilities of derailing one to two railcars but greater probabilities of derailing four to ten railcars, although the difference is very subtle.

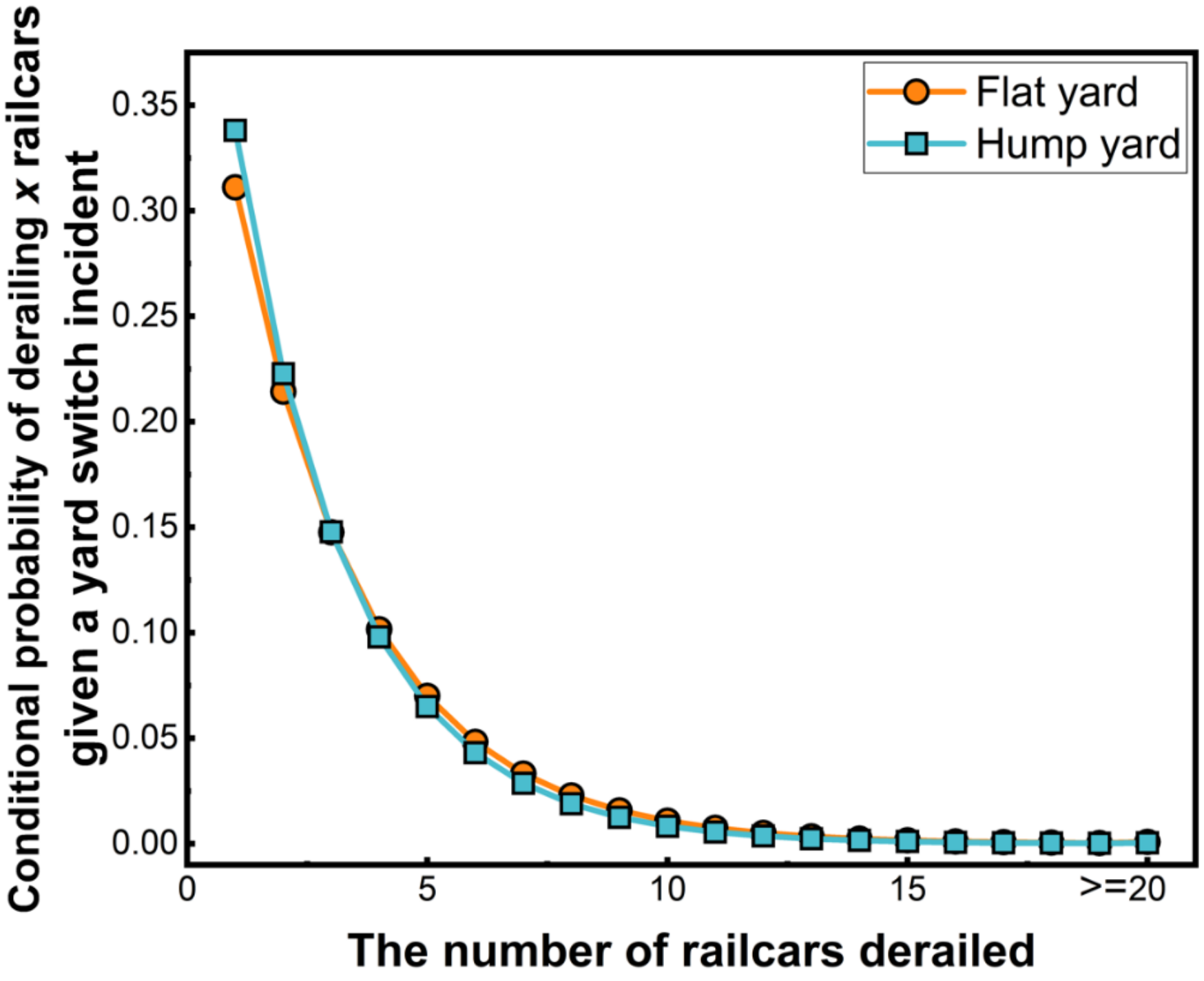

**Figure 8 The conditional probability distribution of the number of railcars derailed given a yard switching incident for different yard types.**

Applying Equations (28) and (29) in the part I paper, given a yard switching incident, the conditional probability distributions for the number of tank cars derailed considering different yard types and switching approaches are shown in Figure 9. When using the "switched alone" approach, the 20 tank cars are assumed to remain in a group and be switched alone, and there is no potential impact from any other non-hazmat railcars derailing in front of and spreading to the tank cars. Given a yard switching derailment when the tank cars are switched alone, there is at least one tank car derailed. Thus, the conditional probability of derailing no tank cars given a yard switching derailment for the switched alone approach is zero for both flat and hump yards. In comparison, when the tank cars are "switched en masse" together with other non-hazmat railcars in front of them, they are exposed to additional risks created if any of the non-hazmat railcars in front of them derail and affect the tank cars. According to empirical data, as discussed in Section 3.3.3 of the part I paper, the assumption is made that yard switching incidents derail a maximum of 20 railcars. As such, to analyze yard switching derailment severity, service options using the "switched en masse" approach assume 19 non-tank cars followed by 20 tank cars. In this case, if the first car of a derailment is the first of the 19 non-tank cars in the group and the derailment spreads back through the railcars to a maximum amount of 20 railcars, the final car to derail will be the first of the 20 tank cars. Considering this approach, there is a possibility that a small derailment starting in the 19 non-hazmat railcars will not be large enough to spread back to the 20 tank cars. For this reason, given a yard switching derailment, the conditional probability of derailing zero tank cars is not zero (0.43 for flat yards and 0.44 for hump yards) when the case study tank cars are "switched en masse" together with non-hazmat cars.

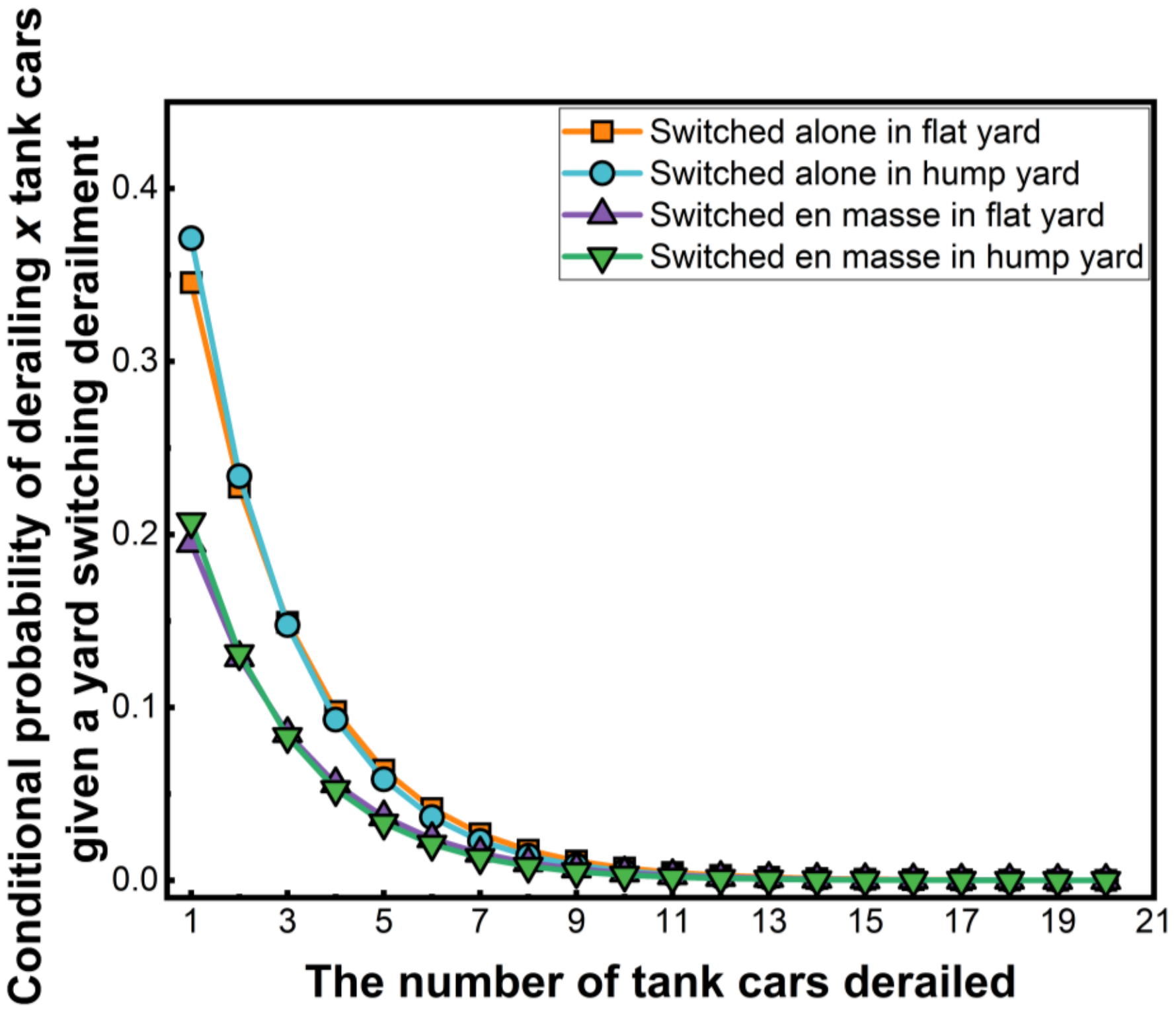

**Figure 9 The conditional probability distributions for the number of tank cars derailed considering different yard types and switching approaches given a yard switching incident.**

Comparing these two yard switching approaches, considering a larger group of cars is switched (39 railcars for the "switched en masse" approach and 20 railcars for the "switched alone" approach), the base likelihood of a yard switching derailment is larger for the "switched en masse" scenarios compared to the "switched alone" scenarios (Table 6). However, many of the yard switching derailments that occur when switching all 39 cars together using the "switched en masse" approach involve mostly non-tank cars or relatively few tank cars. Therefore, in Figure 9, the orange and blue lines hang above the green and purple lines. Although the "switched en masse" approach has a smaller probability of derailing one to two tank cars compared with the "switched alone" approach, the derailment likelihood of the former approach is twice as much as the latter (Table 5). Thus, the "switched en masse" approach generates the "worst-case" scenario regarding yard switching events, while the "switched alone" approach is regarded as the "best-case" scenario.

Applying Equation (26) of the part I paper with yard switching inputs, the probability distribution of the number of tank cars releasing contents given a yard switching derailment is depicted in Figure 10. Since the conditional probability of release for the yard switching event is relatively low ($0.043 \times 0.35 = 0.015$), the probability distributions in Figure 10 skew toward the bottom left corner.

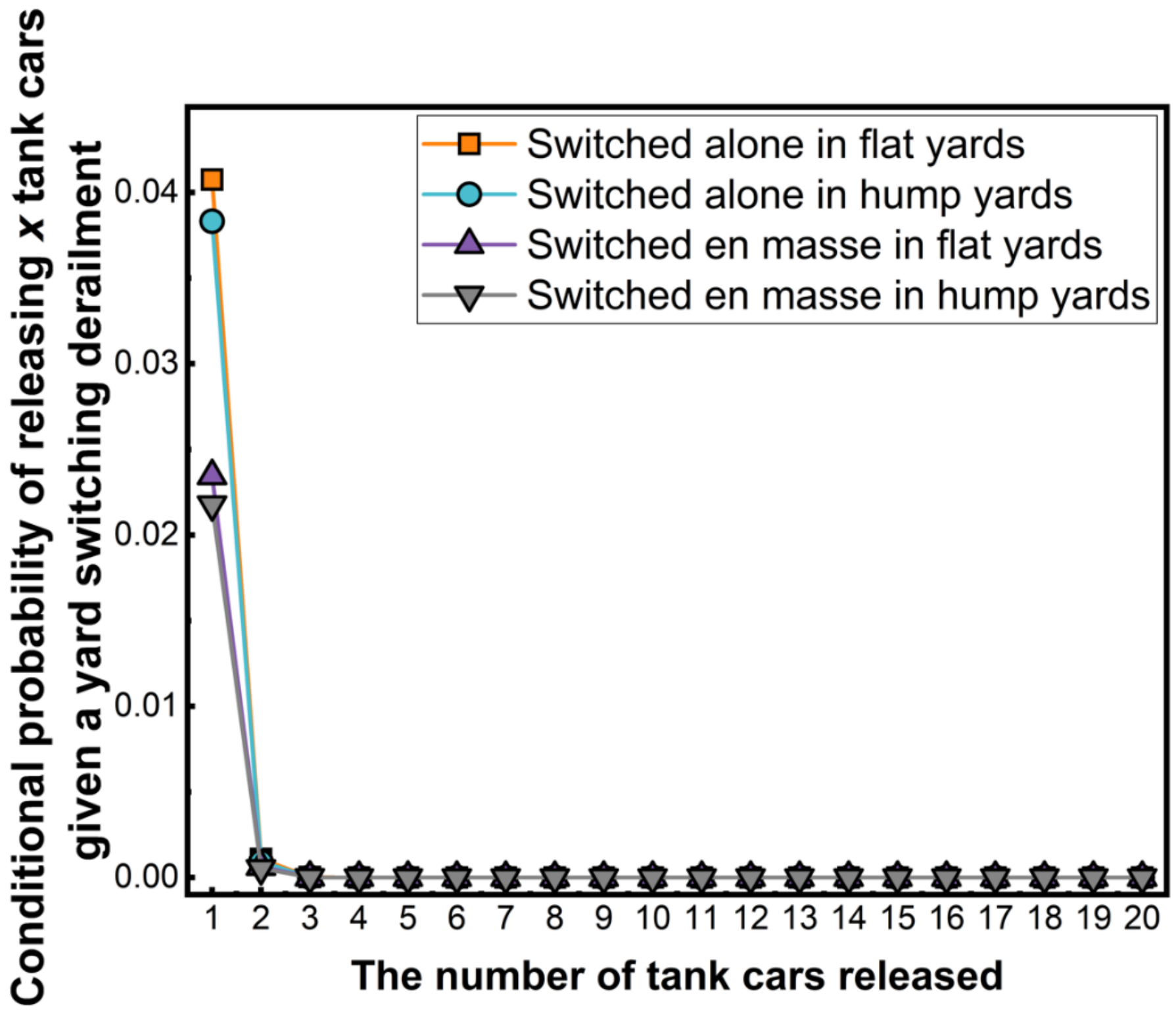

**Figure 10 Conditional probability distribution of the number of tank cars released considering different yard types and switching approaches given a yard switching derailment.**

### 2.3 Total Expected Casualties Per Shipment

The likelihood of a train derailment per shipment considering different risk components, train types, yard types, and yard switching approaches are calculated in Section 2.1. In addition, Section 2.2 determined the conditional probability of releasing a certain number of tank cars per train derailment. Based on these results, the reverse cumulative distributions of the amount released per shipment are calculated according to Sections 3.4 and 3.5 of the part I paper. The reverse cumulative distribution of the amount released details the probability distribution of releasing, in total, more than a certain number of gallons of lading contents. Figure 11 displays the reverse cumulative distribution of the unit train and the manifest train on the mainline segment $i$ (one mile). For the manifest train, the positions of tank cars play an important role in the total amount released: placing tank cars in the middle of the train has a greater probability of releasing a certain amount of lading content compared to placing tank cars at the back of the train (almost two times). Compared with the manifest train, the unit train tends to have a greater probability of releasing a certain amount of lading content on a mainline segment, no matter where the tank cars are placed in a manifest train.

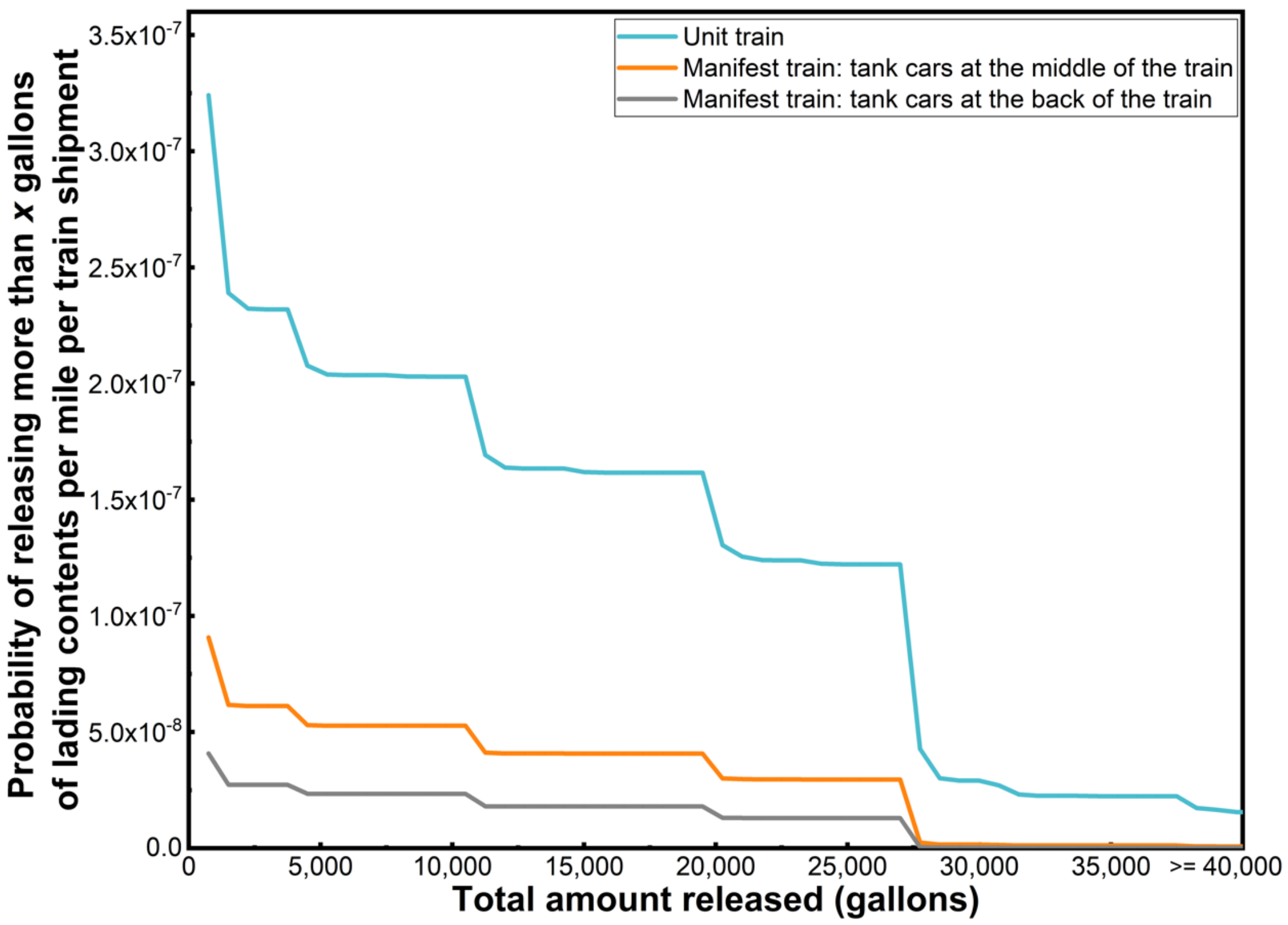

**Figure 11 The reverse cumulative distribution of the amount released per mile per shipment on the mainline segment.**

Figure 12 shows the probability distribution of the amount released for the unit train in all terminals or for the manifest trains in all classification yards for the duration of one shipment. Although the unit train does not have yard switching risks, its probability distribution of releasing a certain amount of lading content during A/D events hangs above most other service options with manifest trains (Figure 12). For low severity releasing incidents (releasing less than 30,000 gallons of lading contents) using manifest trains, generally, flat yards and the "switched en masse" approach exhibit higher risks compared with hump yards and the "switched alone" approach, considering the combined A/D and yard switching risk. To compare the large-scale accidents of the greatest interest in hazmat transportation risk analysis, the reverse cumulative distribution of the amount released overlaps for unit trains in terminals and manifest trains in yards when focusing on releasing more than 30,000 gallons of lading contents. Note that the reverse cumulative distributions on the mainline segment (Figure 11) are approximately two orders of magnitude less than in terminals or yards (Figure 12). This is because the metric on mainline segments is "per shipment per mile" (Figure 11) while in terminals and yards, the metric is "per shipment (considering all terminals and yards encountered)" (Figure 12).

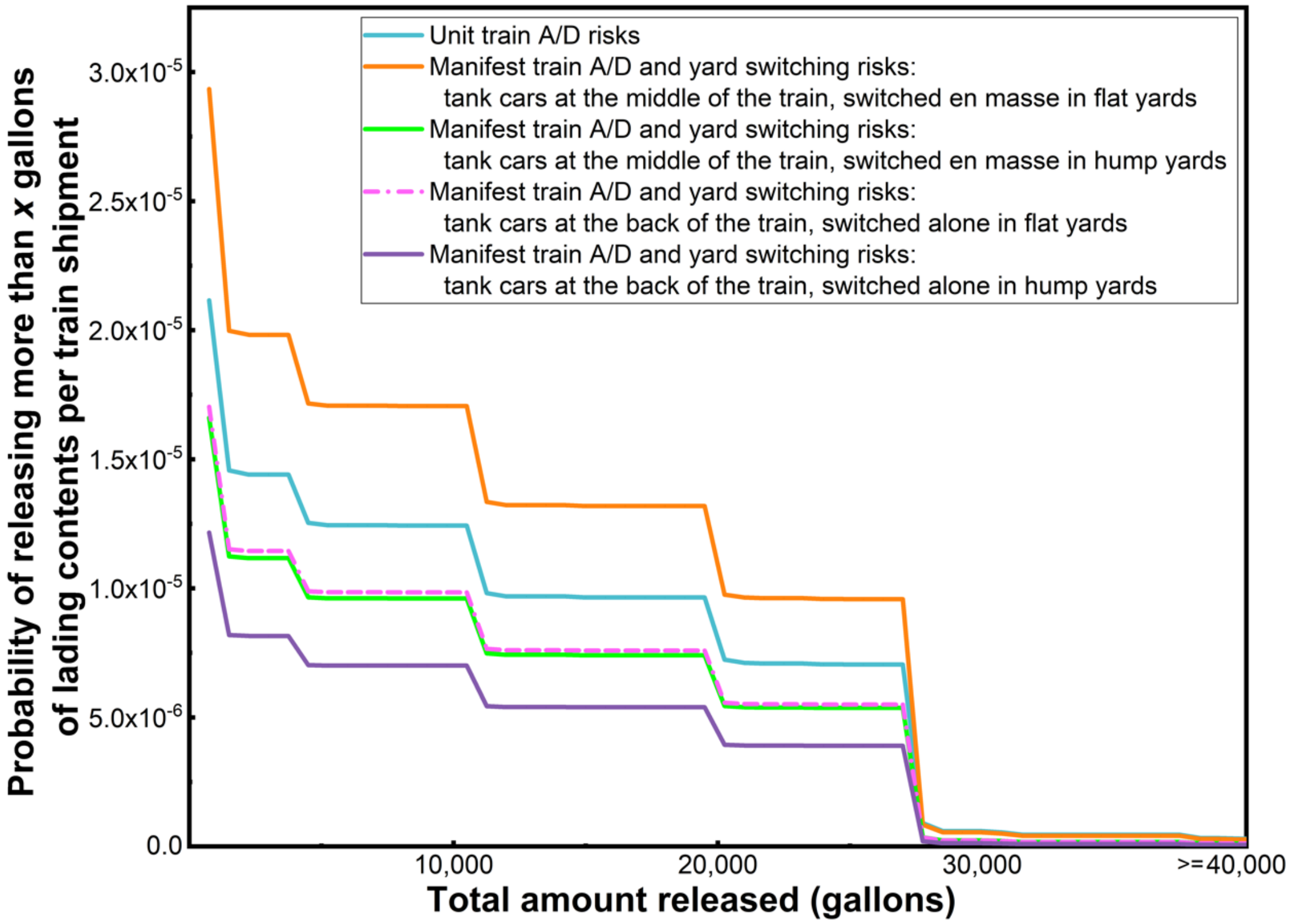

**Figure 12 The reverse cumulative distribution of the amount released per shipment in terminals (for the unit train) or in yards (for the manifest train).**

This paper assumes that the evacuation arrives two hours after the derailment to represent the worst situation. Using the probability distribution of the amount released, constructed above, and the methodology of releasing consequences described in Section 3.6 of the part I paper, the total expected casualties per traffic demand (one shipment for the unit train and five shipments for the manifest train over 400 miles) is summarized in Table 7.

**Table 7 The ranking of the five scenarios concerning the total expected casualties per traffic demand.**

| Ranking (ascending) | Scenario and code | Expected casualties | | | |
|---|---|---|---|---|---|
| | | Line-haul risks | Arrival/departure risks | Yard switching risks | Total per traffic demand |
| 1 | 3 MBAH | 1.04E-04 | 7.80E-05 | | 1.82E-04 |
| 2 | 2 MBAF | 1.04E-04 | 1.10E-04 | | 2.14E-04 |
| 3 | 1 U-T | 1.94E-04 | 2.81E-05 | - | 2.22E-04 |
| 4 | 5 MMEH | 2.36E-04 | 1.07E-04 | | 3.44E-04 |
| 5 | 4 MMEF | 2.36E-04 | 1.91E-04 | | 4.28E-04 |

**Code meaning:**

**U-T: U**nit in **T**erminal.

**MBAF**: **M**anifest, **B**ack of train, switched **A**lone, **F**lat yard type.

**MBAH**: **M**anifest, **B**ack of train, switched **A**lone, and **H**ump yard type.

**MMEF**: **M**anifest, **M**iddle of train, switched **E**n masse, **F**lat yard type.

**MMEH**: **M**anifest, **M**iddle of train, switched **E**n masse, and **H**ump yard type.

The resulting total expected casualties per traffic demand using manifest trains can be compared to using unit trains. Overall, for the case study scenarios, manifest trains with tank cars in the middle of the train and switching the tank cars “en masse” with other railcars in flat yards (scenario 4) have the largest total expected casualties, followed by hump yards with tank cars in the middle and “switched en masse” (scenario 5). This result indicates that placing the tank cars at positions with the highest probability of derailing and using the “switched en masse” approach contributes the most to the overall risks. The service option using one unit train carrying all 100 tank cars in one shipment (scenario 1) ranks third among all scenarios, followed by service options using manifest trains with tank cars at positions with the lowest probability of derailing and “switched alone” (scenarios 2 and 3) in classification yards. Scenario 1 tends to produce results that fall midway between scenarios 2 and 3 and scenarios 4 and 5, reflecting the importance of the tank car positions and switching approaches for manifest trains.

Comparing manifest train service options (comparing scenarios 2 and 4 and scenarios 3 and 5) reveals that routing the case study train through flat yards will exhibit higher expected casualties than routing the case study train through hump yards due to distinct operating strategies, devices, and infrastructure. Changing the position of tank cars in manifest trains and the yard switching approaches (comparing scenarios 2 and 3 with scenarios 4 and 5) can reduce expected casualties by 50% (for flat yards) and 47% (for hump yards).

### 3. Sensitivity Analysis by Train Speed

The case study described and calculated above compares the operating strategies using one unit train or multiple manifest trains transporting 100 hazmat cars over 400 miles. The five factors considered include train type (unit train or manifest train), position in manifest train (at the middle or back of a manifest train), yard switching approach (switched alone or switched en masse), and yard type (flat yard or hump yard). All five scenarios are designed with the assumption that the train derailment speed is 25 mph on the mainline. The fundamental operations in the yard proceed with a reduced speed on non-mainline tracks and yards/terminals. However, the speed typically varies from 25 mph to 50 mph for mainline operations. Thus, the sensitivity analysis is conducted for the overall risks at different speeds. The operating speeds on the mainline are set to 25 mph, 40 mph, and 50 mph for comparative purposes. All other factors remain the same. The operation speed for yards and terminals is still assumed to be 15 mph, due to the operating characteristics of railroad yards and terminals. Table 8 compares expected casualties on the mainline for different derailment speeds. The expected casualties increase with increased derailment speeds.

**Table 8 Expected casualties on mainline for different derailment speeds.**

| Strategies | Derailment speed on the mainline | | |
|---|---|---|---|
| | 25 mph | 40 mph | 50 mph |
| One unit train | 1.94E-04 | 3.95E-04 | 6.79E-04 |
| The best-case train configuration by five manifest trains | 1.04E-04 | 2.26E-04 | 4.37E-04 |
| The worst-case train configuration by five manifest trains | 2.36E-04 | 4.94E-04 | 8.98E-04 |

Table 9 presents the total expected casualties considering mainline risks and yard/terminal risks for various operating speeds on the mainline. The results in Table 9 show that the total expected casualties increase as operating speed increases but changing the operating speed does not change the rank of each scenario as expected. However, a higher speed for the low-rank scenario (e.g., scenario 3 at 50 mph) may have higher expected casualties than a high-rank scenario with a lower speed (e.g., scenario 4 at 25 mph). This indicates that speed plays a vital role in increasing or reducing the expected casualties. At a higher speed, the probability of derailment at each position increases. Once a derailment occurs, it tends to derail more tank cars than the scenarios at lower speeds.

**Table 9 Total expected casualties combining mainline risk and yard risk with various operating speeds on the mainline**

| Scenario and code | Derailment Speed on Mainline | | | Rank (expected casualties low to high) |
|---|---|---|---|---|
| | 25 mph | 40 mph | 50 mph | |
| 3 | 1.82E-04 | 3.04E-04 | 5.15E-04 | 1 |
| MBAH | | | | |
| 2<br>MBAF | 2.14E-04 | 3.36E-04 | 5.47E-04 | 2 |
| 1<br>U-T | 2.22E-04 | 4.24E-04 | 7.07E-04 | 3 |
| 5<br>MMEH | 3.44E-04 | 6.01E-04 | 1.01E-03 | 4 |
| 4<br>MMEF | 4.28E-04 | 6.85E-04 | 1.09E-03 | 5 |

## 4. Insights and Discussions

This paper implements the proposed methodology in the part I paper and designs five scenarios consisting of various levels from train type, yard type, yard switching approach, and tank car placement on manifest trains. Several insights can be concluded from the comparisons among these scenarios.

1) The placement of tank cars in each manifest train and yard switching approach significantly affect the expected total risks. Assume all other contexts are the same; placing tank cars at the lowest-risk positions on a manifest train and switching tank cars "alone" can reduce half of the total risks compared with putting them at the highest-risk positions and switching tank cars "en masse."

2) Although manifest trains experience additional train shipments and additional risks switching and sorting in classification yards, the total transportation risks would also depend on tank car placement in a train, the number of yards on the route, length of the route, and other operational circumstances. Given a certain amount of hazmat to transport, arranging service options should consider safety concerns, economic effects, and operational difficulties. The caveat is that this paper only considers the train configuration problem from safety perspective.

3) Previous paper generally assumed that A/D and yard switching risks in yards/terminals are small since they are operated at a relatively low speed. This part II paper finds that for the unit train operation, it has a comparatively sizeable line-haul risk compared with the A/D risks (the former is almost seven times larger than the latter). However, for manifest train operation, there could be similar magnitude of risks on mainlines and in yards, given specified operating circumstances. This indicates that applying the proposed methodology in part I of this paper could be important (especially accounting for both mainline and yard risks) when comparing unit trains or manifest trains for transporting hazardous materials.

## 5. Concluding Remarks

This paper conducts a case study with five scenarios considering influencing factors of train configuration, the placement of the block of tank cars in a manifest train, yard type, and yard switching approach. The calculation results show that the position of the block of tank cars and the yard switching approach used for manifest trains affect the overall transportation risks significantly. Manifest trains routing through flat yards generally generates higher risks than through hump yards. Specifically, a unit train could have a higher risk than multiple manifest trains when transporting the same amount of hazmat if all tank cars are located at the lowest-risk positions in manifest trains and "switched alone" in classification yards. However, the risk would be higher for service options with multiple manifest trains than for a unit train if all those tank cars were placed at the highest-risk positions and "switched en masse." According to the results from the sensitivity analysis on derailment speeds, a higher speed results in higher potential risks given all else being equal.

Due to data limitations, this paper assumes that the conditional probability of releasing a derailed tank car is the same given the same design and accident speed, and it also assumes that the release quantity of a tank car is independent of other tank cars. These two assumptions are made due to data or information limitations. Given more detailed data used to develop the conditional probability of release in the future, a more accurate conditional probability of release could be built for each position in a train. Furthermore, sensitivity analysis can also be done for different train lengths/number of classification yards encountered for particular interests. Future research can investigate these remaining problems.

## Acknowledgement

This research was funded through a contract by the Federal Railroad Administration (693JJ619C000017). However, all views, analyses, and errors are solely of the authors.

## Appendix A

**Table A.1 Train derailment probability by cause and train type**

**(a) Unit train**

| Derailment cause | Traffic metric used | Derailment probability (one train shipment on a 1-mile segment) |
|---|---|---|
| Broken Rails or Welds | Car mile | 1.53E-07 |
| Broken Wheels (Car) | Car mile | 7.98E-08 |
| Bearing Failure (Car) | Car mile | 6.32E-08 |
| Buckled Track | Train mile | 5.25E-08 |
| Other Axle/Journal Defects (Car) | Car mile | 5.28E-08 |
| Track Geometry (excl. Wide Gauge) | Train mile | 4.87E-08 |
| Obstructions | Train mile | 3.38E-08 |
| Wide Gauge | Train mile | 3.00E-08 |
| Roadbed Defects | Train mile | 2.45E-08 |
| Other Wheel Defects (Car) | Car mile | 2.43E-08 |
| Turnout Defects - Switches | Car mile | 2.26E-08 |
| Track-Train Interaction | Car mile | 2.01E-08 |
| Other Miscellaneous | Train mile | 1.93E-08 |
| Misc. Track and Structure Defects | Train mile | 1.73E-08 |
| Lading Problems | Car mile | 1.60E-08 |
| Joint Bar Defects | Car mile | 1.60E-08 |
| Coupler Defects (Car) | Car mile | 1.42E-08 |
| Other Rail and Joint Defects | Car mile | 1.39E-08 |
| Use of Switches | Train mile | 1.31E-08 |
| Sidebearing, Suspension Defects (Car) | Car mile | 1.25E-08 |
| Train Handling (excl. Brakes) | Train mile | 1.10E-08 |
| Non-Traffic, Weather Causes | Train mile | 1.07E-08 |
| Rail Defects at Bolted Joint | Car mile | 1.04E-08 |
| Train Speed | Train mile | 9.67E-09 |
| Truck Structure Defects (Car) | Car mile | 9.37E-09 |
| Centerplate/Carbody Defects (Car) | Car mile | 7.64E-09 |
| All Other Car Defects | Train mile | 7.60E-09 |
| Misc. Human Factors | Train mile | 7.25E-09 |
| Stiff Truck (Car) | Train mile | 5.18E-09 |
| Switching Rules | Train mile | 5.18E-09 |
| Failure to Obey/Display Signals | Train mile | 4.83E-09 |
| Other Brake Defect (Car) | Car mile | 4.86E-09 |
| Handbrake Operations | Train mile | 4.14E-09 |
| Brake Rigging Defect (Car) | Car mile | 4.17E-09 |
| Loco Electrical and Fires | Train mile | 3.80E-09 |
| Track/Train Interaction (Hunting) (Car) | Car mile | 3.47E-09 |
| Brake Operation (Main Line) | Car mile | 3.12E-09 |
| Mainline Rules | Train mile | 3.11E-09 |
| Signal Failures | Car mile | 2.78E-09 |
| Loco Trucks/Bearings/Wheels | Car mile | 2.78E-09 |
| Turnout Defects - Frogs | Car mile | 1.74E-09 |
| All Other Locomotive Defects | Train mile | 1.04E-09 |
| Brake Operations (Other) | Train mile | 6.90E-10 |
| UDE (Car or Loco) | Car mile | 3.47E-10 |
| Employee Physical Condition | Train mile | 3.45E-10 |
| Air Hose Defect (Car) | Car mile | 3.47E-10 |
| Total derailment probability | | 8.53E-07 |

**(b) Manifest train**

| Derailment cause | Traffic metric used | Derailment probability (one train shipment on a 1-mile segment) |
|---|---|---|
| Broken Rails or Welds | Car mile | 1.39E-07 |
| Track Geometry (excl. Wide Gauge) | Train mile | 4.75E-08 |
| Bearing Failure (Car) | Car mile | 7.44E-08 |
| Train Handling (excl. Brakes) | Train mile | 3.94E-08 |
| Obstructions | Train mile | 2.95E-08 |
| Track-Train Interaction | Car mile | 4.60E-08 |
| Lading Problems | Car mile | 4.58E-08 |
| Wide Gauge | Train mile | 2.26E-08 |
| Coupler Defects (Car) | Car mile | 3.99E-08 |
| Use of Switches | Train mile | 2.21E-08 |
| Broken Wheels (Car) | Car mile | 3.75E-08 |
| Sidebearing, Suspension Defects (Car) | Car mile | 3.56E-08 |
| Other Wheel Defects (Car) | Car mile | 3.56E-08 |
| Brake Operation (Main Line) | Car mile | 3.54E-08 |
| Centerplate/Carbody Defects (Car) | Car mile | 3.21E-08 |
| Buckled Track | Train mile | 1.79E-08 |
| Other Miscellaneous | Train mile | 1.76E-08 |
| Turnout Defects - Switches | Train mile | 1.73E-08 |
| Misc. Track and Structure Defects | Train mile | 1.19E-08 |
| Train Speed | Train mile | 1.14E-08 |
| Stiff Truck (Car) | Train mile | 1.03E-08 |
| Roadbed Defects | Train mile | 9.96E-09 |
| Joint Bar Defects | Car mile | 1.52E-08 |
| Other Axle/Journal Defects (Car) | Car mile | 1.39E-08 |
| Other Brake Defect (Car) | Car mile | 1.39E-08 |
| Loco Trucks/Bearings/Wheels | Car mile | 1.37E-08 |
| All Other Car Defects | Train mile | 7.53E-09 |
| Track/Train Interaction (Hunting) (Car) | Car mile | 1.26E-08 |
| Misc. Human Factors | Train mile | 7.05E-09 |
| Switching Rules | Train mile | 6.68E-09 |
| Other Rail and Joint Defects | Car mile | 1.11E-08 |
| Rail Defects at Bolted Joint | Car mile | 1.11E-08 |
| Handbrake Operations | Train mile | 5.95E-09 |
| Non-Traffic, Weather Causes | Train mile | 5.35E-09 |
| Failure to Obey/Display Signals | Train mile | 4.74E-09 |
| Brake Rigging Defect (Car) | Car mile | 7.59E-09 |
| All Other Locomotive Defects | Train mile | 4.25E-09 |
| Signal Failures | Car mile | 7.59E-09 |
| Air Hose Defect (Car) | Car mile | 7.16E-09 |
| Truck Structure Defects (Car) | Car mile | 5.42E-09 |
| Loco Electrical and Fires | Train mile | 2.79E-09 |
| Mainline Rules | Train mile | 2.79E-09 |
| Turnout Defects - Frogs | Car mile | 4.34E-09 |
| Radio Communications Error | Train mile | 1.46E-09 |
| UDE (Car or Loco) | Car mile | 2.17E-09 |
| Brake Operations (Other) | Train mile | 7.29E-10 |
| TOFC/COFC Defects | Train mile | 6.08E-10 |
| Employee Physical Condition | Train mile | 2.43E-10 |
| Handbrake Defects (Car) | Train mile | 2.43E-10 |
| Total derailment probability | | 9.54E-07 |